\documentclass[%
reprint,                      
superscriptaddress,
preprintnumbers,              
amsmath,amssymb,
aps,
prb,
floatfix,
]{revtex4-1}

\usepackage{graphicx}
\usepackage{dcolumn}
\usepackage{bm}
\usepackage[mathlines]{lineno} 

\newcommand{\vm}[1]{\mathrm{#1}}

\begin{document}

\preprint{revision 2.718281828459045}

\title{First principles calculations of anisotropic charge carrier mobilities
  in organic semiconductor crystals} 

\author{V. Stehr}
\email{Stehr@Physik.Uni-Wuerzburg.de}
\affiliation{Physikalisches Institut,
  Universit\"at W\"urzburg, 97074 W\"urzburg, Germany}
\author{J. Pfister}
\author{R. F. Fink}
\affiliation{Institut f\"ur Physikalische und Theoretische Chemie,
  Universit\"at W\"urzburg, 97074 W\"urzburg, Germany}
\author{B. Engels}
\affiliation{Institut f\"ur Physikalische und Theoretische Chemie,
  Universit\"at W\"urzburg, 97074 W\"urzburg, Germany}
\author{C. Deibel}
\affiliation{Physikalisches Institut,
  Universit\"at W\"urzburg, 97074 W\"urzburg, Germany}


\begin{abstract}
The orientational dependence of charge carrier mobilities in organic
semiconductor crystals and the correlation with the crystal structure are
investigated by means of quantum chemical first principles calculations
combined with a model using hopping rates from Marcus theory. A master
equation approach is presented which is numerically more efficient than the
Monte Carlo method frequently applied in this context. Furthermore,
it is shown that the widely used approach to calculate the mobility via the
diffusion constant along with rate equations is not appropriate in many
important cases. The calculations are compared with experimental data, showing
good qualitative agreement for pentacene and rubrene. In addition, charge
transport properties of core-fluorinated perylene bisimides are investigated.
\end{abstract}

\maketitle


\section{Introduction}

Due to their low production costs and easy processability organic
semiconductor devices are promising materials for organic light emitting
diodes (OLEDs),~\cite{armstrong2009,meerheim2009,kulkarni2004} organic field
effect transistors (OFETs),~\cite{dimitrakopoulos2002,newman2004,facchetti2007}
radio frequency identification tags (RFIDs)~\cite{subramanian2005,myny2010}
and solar
cells,~\cite{riede2008,kippelen2009,dennler2009,chidichimo2010,deibel2010} to
mention just a few. The performance of these devices depends crucially on
the charge transport. Therefore, it is important to  understand
the basic principles of charge transport in these materials.

Various models have been proposed which are often contradictory.
The band theory, which is well established for inorganic covalently
bonded materials, is not particularly appropriate for organic conductors,
because organic molecular crystals are only weakly bound by van der Waals
interactions causing the molecules to be much more flexible. Due to the
complex nodal structure of the molecular orbitals the transfer integrals
between the monomers are very sensitive to even small nuclear
displacements. That is why lattice vibrations play a more important role in
organic than in inorganic materials, as they destroy the long range
order and lead to a charge carrier localization.~\cite{bredas2009}
To account for these vibrations, a variety of models have been proposed which
incorporate the local (Holstein)~\cite{holstein1959} and the nonlocal
(Peierls)~\cite{peierls} coupling. The latter leads to a polaron model where
the charge carrier is partially localized and dressed by
phonons.~\cite{hannewald2004,hannewald2004a,ortmann2010,ortmann2010a}
The fluctuations of the coupling between the molecules are of the same order
of magnitude as the average coupling,~\cite{troisi2005} leading to a rather
strong localization. Other models have been suggested, where the charges are
assumed to be localized and the inter- and intramolecular vibrations are
treated classically.~\cite{troisi2006,troisi2007,stafstroem2010}

At higher temperatures, it is often appropriate to assume that the charge is
localized due to the thermal disorder of the molecules and that charge
transport occurs via thermally activated hopping.~\cite{schein1978}
In some cases room
temperature should be sufficient for this assumption to be justified.
We apply this hopping model to study the dependence of the charge carrier mobility on
the molecular structure and morphology as well as its angular dependency. The
latter point is important since most organic crystals show a pronounced
anisotropy for the transport parameters which has to be taken into account for
device design. Furthermore, it is known that the mobility is very sensitive to
the arrangement of the monomers and that already small changes in their
alignment can alter the transport parameters dramatically.~\cite{bredas2002}

A promising class of materials for organic electronics are perylene
bisimides. Due to their light resistance~\cite{herbst} and intense
photoluminescence~\cite{langhals1998} they are widely used as robust organic
dyes in the automobile industry.~\cite{herbst} Furthermore, they show a
considerable electron mobility~\cite{struijk2000,chen2004,an2005} and a high
electron affinity.~\cite{schmidt2007,chen2004} That is why they serve as
n-type semiconductors for organic field effect
transistors~\cite{chen2007,schmidt2007,weitz2008,molinari2009,wen_b_2009,zhan2007}
and as electron acceptor material in organic solar
cells.~\cite{zhan2007,woehrle,schmidt-mende2001,shin2006}

Section~\ref{sec:theory} describes the theoretical background of the applied
model as well as details of the numerical calculations and computational
approaches. It is shown that the master equation approach is particularly faster
than the well-known Monte Carlo method. Furthermore we elucidate why the
commonly applied approach to calculate the mobility via the diffusion constant
along with rate equations~\cite{wen2009,liu2010,deng2004,coropceanu2007} is
not appropriate in many important cases. In Sec.~\ref{sec:einstein} we
consider the frequently disputed question if the Einstein relation holds even
for more disordered (amorphous)
materials.~\cite{harada2005,pautmeier1991,borsenberger1991,casado1994,
baranovskii1998a} In Sec.~\ref{sec:anisotropy} we show results for the
orientational and morphological dependency of the mobility for pentacene,
rubrene and two fluorinated perylene bisimides. The first two materials are
experimentally and theoretically well
investigated~\cite{mattheus2001,malagoli2004,jurchescu2004,lee2006,
coropceanu2007,wen2009,sundar2004,podzorov2004,boer2004,dasilva2005,
jurchescu2006,zeis2006,ling2007}
which allows for the comparison with experimental data.

\section{Theory and modeling} \label{sec:theory}

\subsection{The Marcus hopping model} \label{sec:hopping_model}

In this work, a hopping mechanism is assumed for the motion of the charge
carriers. The hopping rate from a site $i$ to $j$ is given by the Marcus
equation~\cite{marcus1956,marcus1993}
\begin{equation} \label{eq:marcus}
  \nu_{ji} = \frac{{V_{ji}}^2}{\hbar} \sqrt{\frac{\pi}{\lambda k_{\vm{B}} T}}
      \exp \left[ -
        \frac{(\Delta E_{ji} + \lambda)^2}{4 \lambda k_{\vm{B}} T} \right],
\end{equation}
where $V_{ji}$ is the electronic coupling parameter, $\lambda$ is the
reorganization energy, $T$ is the temperature, $k_{\vm{B}}$ is the Boltzmann
constant and $\hbar = h/(2 \pi)$ where $h$ is the Planck constant. The energy
difference $\Delta E_{ji}$ between the two hopping sites is caused by an
external electric field $\vec F$. If the material is less ordered or even
amorphous, each molecule experiences slightly different surrounding effects
(such as polarization) that lead to different site energies $E_i^0$. These
energy differences furthermore contribute to $\Delta E_{ji}$: 
\begin{equation} \label{eq:enthalpy}
  \Delta E_{ji} = (E_j^0 - E_i^0) - q \vec F \vec r_{ji},
 \end{equation}
where $q$ is the charge which equals the positive or negative unit charge
and $\vec r_{ji}$ is the distance vector between sites $i$ and $j$.
Marcus rates have been used before for calculating the anisotropy of the
charge carrier mobility,~\cite{wen2009} but with $\Delta E_{ji} \equiv 0$.

The interaction of the charge carriers with the phonons is partially considered
by the reorganization energy. Due to the weak van der Waals interactions
between organic molecules, it can be divided into an internal (intramolecular)
and an external (intermolecular) part, i.e.\
$\lambda = \lambda_{\vm{int}} + \lambda_{\vm{ext}}$.
The intramolecular reorganization energy $\lambda_{\vm{int}}$ is due to the
geometry changes of the donor and the acceptor monomer upon the charge
transfer process. The external reorganization energy $\lambda_{\vm{ext}}$
covers the energetic changes concerning the surrounding, caused by lattice
distortion and polarization. For oligoacenes $\lambda_{\vm{ext}}$ was shown to
be about one order of magnitude smaller than
$\lambda_{\vm{int}}$.~\cite{norton2008,mcmahon2010} Furthermore, is was 
demonstrated that $\lambda_{\vm{int}}$ of a molecule is lower in
a cluster than in gas phase and that the total reorganization energy of
naphthalene  is closer to $\lambda_{\vm{int}}$ in the gas phase
than to $\lambda_{\vm{int}}$ in the cluster.~\cite{norton2008}
That is why the external reorganization energy is neglected in this paper and
the internal reorganization energy of the monomer in vacuum is used
for~$\lambda$.

The Marcus theory was originally derived for outer sphere electron
transfer in solvents.~\cite{marcus1956} It stems from time dependent
perturbation theory (Fermi's Golden rule) and describes a non-adiabatic charge
transfer where the charge carrier is localized at the donor or acceptor
molecule respectively. Treating the coupling as a perturbation requires
that $V_{ji}$ is small compared to $\lambda/4$, which corresponds to the
activation energy for the charge carrier to change place (for
$\Delta E_{ji} = 0$). Furthermore, the thermal relaxation (the geometric
reorganization) has to be fast in comparison with the transfer so that the
system can be assumed to be in thermal equilibrium during the transfer. In
addition, the theory is restricted to the high temperature case since
tunneling is neglected completely and the molecular vibrations are treated
classically, what requires $k_{\vm{B}}T \gg \hbar \omega$. These restrictions
of the Marcus theory in the context of charge transport are discussed
elsewhere.~\cite{picon2007,cheung2008} Despite all imperfections it is
widely used for charge transfer in organic
crystals~\cite{deng2004,li2007,wen2009,liu2010,sancho-garcia2008,sancho-garcia2010,sancho-garcia2008b}
and one can certainly assume that this theory is suitable for the purpose of a
qualitative charge transport analysis.

\subsection{The master equation approach} \label{sec:meq}

The master equation  approach was used to describe the transport process. In
the case of low charge carrier densities, the master equation, which describes
the hopping of the charge carriers in the organic semiconductor, has the
simple linear form~\cite{houili2006}
\begin{equation} \label{eq:master}
  \frac{d p_i}{d t} = \sum_j (\nu_{ij} p_j - \nu_{ji} p_i),
\end{equation}
where $p_i$ denotes the probability that the lattice site $i$ is occupied by a
charge carrier. The index $j$ sums over all other sites. In principle,
it is also possible to include repulsive forces between the charge carriers in
the master equation in order to account for higher charge carrier
densities. However, in the case of low densities, even  the quite simple
Eq.~(\ref{eq:master}) leads to good results. 

In the steady state, a dynamic balance is reached where the occupation
probabilities for the sites do not change anymore and $dp_i/dt$ in
Eq.~(\ref{eq:master}) equals zero. Since this equation holds for all sites in
the crystal, this results in a linear system of equations,
\begin{equation} \label{eq:matrixequation}
  \mathbf{N} \cdot \vec p = \vec 0.
\end{equation}
$\vec p $ contains the unknown $p_i$ and $\mathbf{N}$ is a negative
semidefinite sparse matrix that contains all hopping rates $\nu_{ji}$.
For one dimension $\mathbf{N}$ is
  \begin{equation} \label{eq:matrix}
    \left(
    \begin{array}{c|cccc|c}
      & \vdots & \vdots & \vdots & \vdots &\\
      \hline
      \ldots & - \sum_j \nu_{j1} & \nu_{12} & \nu_{13} & \nu_{14} & \ldots\\
      \ldots & \nu_{21} & - \sum_j \nu_{j2} & \nu_{23} & \nu_{24} & \ldots\\
      \ldots & \nu_{31} & \nu_{32} & - \sum_j \nu_{j3} & \nu_{34} & \ldots\\
      \ldots & \nu_{41} & \nu_{42} & \nu_{43} & - \sum_j \nu_{j4} & \ldots\\
      \hline
      & \vdots & \vdots & \vdots & \vdots &
    \end{array}
    \right).
  \end{equation}
The columns correspond to the initial sites $i$ of the charge carrier and the
lines correspond to the final sites $j$, i.e., the jump rate $\nu_{ji}$ from
$i$ to $j$ appears in the $i$th column and the $j$th line. The diagonal
elements contain the negative sum of all hopping rates away from the
respective site.

The infinite matrix $\mathbf{N}$ is approximated by a finite matrix with
cyclic boundary conditions, i.e., a charge carrier that leaves the crystal
at one side reenters at the opposite side. This means for the example matrix
depicted in Eq.~(\ref{eq:matrix}) that the charge which jumps from site 4 in
positive direction ends at site 1. For this boundary condition to be
applicable it has to be assured that the hopping rate from site 4 to site 1
in negative direction is negligible. This results in a constraint for the
minimum size of the matrix.

The matrix in Eq.~(\ref{eq:matrix}) was extended to three dimensions resulting
in a $(3n_dn_m) \times (3n_dn_m)$ matrix where $n_d$ is the number of unit
cells in each direction and $n_m$ is the number of monomers per unit cell. In
this work all monomers within a cube of three unit cells length in each
dimension of the crystal are taken into account. It was verified that a bigger
matrix with more than $3 \times 3 \times 3$ unit cells does not change the
result. The hopping rates were calculated from one monomer to all other
monomers in the same and in the adjacent cells. Since the jump rate,
Eq.~(\ref{eq:marcus}), implicitly depends on the distance via the electronic
coupling $V_{ji}$, larger jump distances can be neglected. 

Solving Eq.~(\ref{eq:matrixequation}) and taking into account the
normalization condition $\sum_i p_i = 1$ provides the occupation probabilities
for all sites. (For $\Delta E_{ji} = 0$, it is the same for all sites.)
These probabilities can then be used to calculate the mobility of the charge
carriers in field direction from
\begin{equation} \label{eq:mu-def}
  \mu = \frac{\langle v \rangle}{F},
\end{equation}
with the average velocity
\begin{equation}
  \langle v \rangle = \sum_i p_i v_i
                    = \sum_i p_i \frac{\langle r_{\parallel}  \rangle_i}{\tau_i},
\end{equation}
where $v_i$ is the resulting velocity at site $i$,
\begin{equation}
  \langle r_{\parallel}  \rangle_i 
  = \frac{\sum_j \nu_{ji} \left( \vec r_{ji} \frac{\vec F}{F} \right)}{\sum_j \nu_{ji}}
\end{equation}
is the average displacement at site~$i$ in field direction and
\begin{equation} \label{eq:dwelltime}
  \tau_i = \left( \sum_j \nu_{ji} \right)^{-1}
\end{equation}
is the dwell time of the charge carrier at site~$i$.
Equations~(\ref{eq:mu-def}) to (\ref{eq:dwelltime}) result in~\cite{yu2001}
\begin{eqnarray}
  \mu &=& \frac{1}{F} \sum_i \left( p_i \sum_j \nu_{ji} \frac{\sum_j \nu_{ji}
      \left( \vec r_{ji} \frac{\vec F}{F} \right)}{\sum_j \nu_{ji}} \right) \nonumber \\
      &=& \frac{1}{F} \sum_{ij} p_i \nu_{ji} \vec r_{ji} \frac{\vec F}{F} \label{eq:mob1}.
\end{eqnarray}

In order to simplify the calculation of the mobility within such a jump rate
approach, the mobility is often calculated without external field because the
occupation probabilities of the sites do not differ in this case and one does
not have to solve the master equation~(\ref{eq:matrixequation}). Since
Eq.~(\ref{eq:mob1}) is not applicable in that case (because $F = 0$), the
mobility is calculated via the diffusion coefficient~$D$ and the Einstein
relation~\cite{landsberg1981}
\begin{equation} \label{eq:einstein}
  \mu = \frac{q}{k_{\vm{B}} T} D.
\end{equation}

Different equations are found in the
literature~\cite{wen2009,deng2004,liu2010,coropceanu2007,bisquert2008} to
evaluate $D$. Considerations similar to those above for the mobility seem to
provide
\begin{equation} \label{eq:D-def}
  D = \frac{1}{2n} \frac{d}{dt} \langle r^2 \rangle
    = \frac{1}{2n} \sum_i p_i \frac{\langle r^2 \rangle_i}{\tau_i},
\end{equation}
where $n$ is the spatial dimensionality. Since the diffusion is regarded in
one dimension here, $n$ equals 1 and
\begin{equation}
  D = \frac{1}{2} \sum_i p_i \frac{\langle r^2_{\parallel} \rangle_i}{\tau_i},
\end{equation}
where
\begin{equation} \label{eq:variance}
  \langle r^2_{\parallel} \rangle_i = \frac{\sum_j \nu_{ji} (\vec r_{ji} \vec e)^2}{\sum_j \nu_{ji}}
\end{equation}
is the variance of the charge carrier position at site~$i$ in the
direction of the unit vector~$\vec e$. Equations~(\ref{eq:dwelltime})
and~(\ref{eq:D-def}) to~(\ref{eq:variance}) finally result in
\begin{equation} \label{eq:D}
  D = \frac{1}{2} \sum_{ij} p_i \nu_{ji}
  \left( \vec r_{ji} \vec{e} \right)^2
\end{equation}
for the diffusion coefficient in the direction of $\vec{e}$. It is worth
mentioning that Eq.~(\ref{eq:D}) holds even in the presence of an external
field (see Appendix~\ref{sec:appendix}).

Without external field and assuming that all lattice
sites are equal (i.e.\ $\Delta E_{ji} =0$), the last equation simplifies
to~\cite{coropceanu2007,zhdanov1985,myshlyavtsev1995}
\begin{equation} \label{eq:D-simple}
  D = \frac{1}{2} \sum_j \nu_j (\vec r_j \vec{e})^2.
\end{equation}

It is important to note, that the diffusion constants in Eqs.~(\ref{eq:D})
and~(\ref{eq:D-simple}) are not strictly correct. Just if the unit cell of the
crystal contains only a single molecule and if the crystal structure is
perfectly translation-symmetric, i.e.\ $E_i^0 = E_j^0$ for all monomer pairs,
cf.\ Eq.~(\ref{eq:enthalpy}), these equations become correct.

\begin{figure}
  \includegraphics[width=6cm]{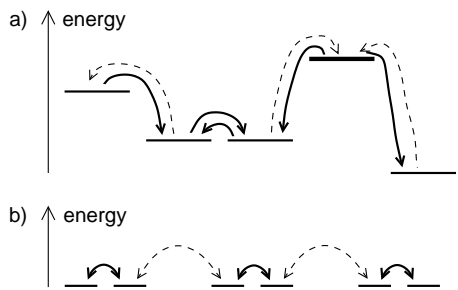}
  \caption{\label{fig:trapping} The charge carrier is ``trapped'' between two
    lattice sites. a) The surrounding of the monomer causes an energetic
    ``pit''. b) Strongly differing jump rates lead to a capturing.}
\end{figure}

However, in less ordered or even amorphous materials the site
energies~$E_{i}^0$ and~$E_{j}^0$ are different
because of the differing surroundings for each lattice site. In that case,
the occupation probabilities~$p_i$ differ and the master equation has to be
applied. In the case of strongly different $E_{i}^0$, even Eq.~(\ref{eq:D})
becomes incorrect since the charge carrier can be ``trapped'' between two
lattice sites with similar energy,~\cite{gonzalez-vazquez2009} see
Fig.~\ref{fig:trapping}a: Because of the energetically unfavorable
surrounding, the charge carrier jumps back and forth between the same sites
all the time. These moves do not contribute to a macroscopic spreading of the
occupation probability of the charge carrier with the time. That is why the
averaging in Eq.~(\ref{eq:D}) overestimates the true macroscopic diffusion
coefficient. This problem does not appear in Eq.~(\ref{eq:mob1}) since the
$\vec r_{ji}$ is not squared as in Eq.~(\ref{eq:D}). For that reason the
contribution of the trapped charge cancels when summing over all lattice
sites. And even in perfectly ordered crystals where all jump rates are
symmetric, i.e.\ $\nu_{ji} = \nu_{ij}$ (without external field), such a
trapping can occur if different sites exist in the elementary cell of the
crystal and if the hopping rates within the cells differ from those to
neighbored unit cells, see Fig.~\ref{fig:trapping}b: Here, the charge carrier
jumps back and forth between two monomers with a high coupling because the
coupling to the other neighbors is lower. In such cases Eq.~(\ref{eq:mob1}) in
conjunction with Eq.~(\ref{eq:einstein}) provides correct diffusion
coefficients while Eqs.~(\ref{eq:D}) and~(\ref{eq:D-simple}) overestimate the
values for $D$.

\subsection{The Monte Carlo approach} \label{sec:montecarlo}

The master equation results were verified with Monte Carlo simulations
applying the algorithm of Houili {\it et al.},~\cite{houili2006} but
without any interaction between the charge carriers.
The mobility and the diffusion coefficient were calculated via
\begin{equation} \label{eq:mu-mc}
  \mu = \frac{1}{F} \frac{d}{d t}  \left \langle \vec{r}_{ji}
    \frac{\vec F}{F} \right \rangle
\end{equation}
and
\begin{equation} \label{eq:D-mc}
  D = \frac{1}{2} \frac{d}{d t} \left \langle \left(\vec{r}_{ji}
    \vec e - \left \langle \vec{r}_{ji} \vec e \right \rangle \right)^2 \right
    \rangle 
\end{equation}
respectively. The time dependent average position
$\langle \vec{r}_{ji} \frac{\vec F}{F} \rangle$ and the variance
$\langle \left(\vec{r}_{ji} \vec e - \left \langle \vec{r}_{ji} \vec e \right
  \rangle \right)^2 \rangle$
have been averaged  over a sufficient number of simulation runs to obtain
smooth lines. It was checked that both average and variance show a linear time
dependence in order to secure the stationary state.

The Monte Carlo approach is just an alternative way to solve the master
equation~(\ref{eq:master}). It is a feasible way to log the atomic scale
motions underlying the transport properties as a function of time.
However, as this is a stochastic method, many simulation runs are needed in
order to achieve an acceptably low statistical error such that sufficiently
significant values are obtained for the mobility and the diffusion
coefficient. Furthermore, one has to take care that the stationary state is
reached within the simulation time. This is a serious problem in the case of
strongly disordered materials. In contrast to that, the approach used here by
solving the matrix equation~(\ref{eq:matrixequation}) which provides the
stationary state by means of analytic numerical methods guarantees the
stationary solution and is  furthermore numerically more efficient than Monte
Carlo simulations.~\cite{yu2001}

\subsection{Quantum chemical methods} \label{sec:qc}

\begin{figure}
  \includegraphics[width=5cm]{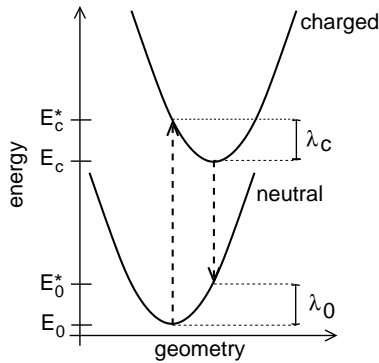}
  \caption{\label{fig:reorganisation} The potential energy surfaces of the
    neutral and the charged monomer. The dashed arrows indicate the vertical
    transitions from one state to the other. $\lambda_0$ and $\lambda_c$ are
    the two contributions to the reorganization energy,
    see Eq.~(\ref{eq:lambda}).}
\end{figure}

The electronic coupling~$V_{ji}$ and the reorganization energy~$\lambda$
needed for the hopping rate, Eq.~(\ref{eq:marcus}), are determined by quantum
chemical first principles calculations. In order to calculate $\lambda$,
the geometry of the isolated monomer was optimized for the charged and
the neutral state. The energies $E_0$ and $E_c$ of the neutral and the charged
monomers in their lowest energy geometries and the energies $E_0^*$ and
$E_c^*$ of the neutral monomer with the ion geometry and the charged monomer
with the geometry of the neutral state are calculated to get the intramolecular
reorganization energy~\cite{malagoli2004}
\begin{equation} \label{eq:lambda}
  \lambda = \lambda_c + \lambda_0 = (E_c^* - E_c) + (E_0^* - E_0),
\end{equation}
cf.\ Fig.~\ref{fig:reorganisation}. For all quantum chemical calculations
the TURBOMOLE program package~\cite{turbomole60} was used.
The calculations were conducted via  density functional theory using the
hybrid generalized gradient functional
B3-LYP~\cite{dirac1929,slater1951,vosko1980,becke1988,lee1988,becke1993}
with the correlation consistent polarized valence double zeta basis set
(cc-pVDZ)~\cite{dunning1989} for all atoms. This functional was chosen because
it has been shown that it leads to quite good results for describing the
ionization-induced geometry modifications of
oligoacenes.~\cite{sanchez-carrera2006,coropceanu2002}

The electronic couplings were calculated as described by
Li {\it et al.}~\cite{li2007} resulting in
\begin{equation}
  V_{ji} = \frac{H_{ji} - \frac{1}{2} (H_{ii} + H_{jj}) S_{ji}}{1 - S_{ji}^2}
\end{equation}
with
\begin{eqnarray}
  H_{ji} &=& \langle \varphi_j | \hat H_{KS} | \varphi_i \rangle \nonumber\\
  S_{ji} &=& \langle \varphi_j | \varphi_i \rangle. \nonumber
\end{eqnarray}
For hole (electron) transport $\varphi_i$ and $\varphi_j$ are the HOMO (LUMO)
orbitals of the respective isolated monomers and $\hat H_{KS}$ is the
Kohn-Sham operator of the neutral dimer system. $H_{ii}$ and $H_{jj}$ are the
site energies of the two monomers, $S_{ji}$ is the spatial overlap and
$H_{ji}$ is the charge transfer integral in the non-orthogonalized basis.

The arrangement of the monomers in the crystal was extracted from X-ray crystal
structure data which was retrieved from the Cambridge Structural Database.

\subsection{The Gaussian disorder model} \label{sec:disorder}

It has been argued that the Einstein relation, Eq.~(\ref{eq:einstein}), does
not hold in disordered organic
materials in general~\cite{pautmeier1991,borsenberger1991,casado1994}
or at least if additionally an external field is
applied.~\cite{bisquert2008,anta2008,novikov2006a} In fact it turned out that
this is only true for rather high charge carrier densities,~\cite{roichman2002}
low temperatures and high electric fields which are out of the scope of the
present work. At extremely low temperatures, the thermal energy of the charge
carriers is not sufficient to reach sites which are higher in energy and only
energy-loss jumps occur. In that case, neither $\mu$ nor $D$ depends on the
temperature.~\cite{baranovskii1998} For low fields, the transport coefficients
are independent of the field,~\cite{abkowitz1991,baessler1993}
but for higher fields nonlinear effects become important and $D/\mu$ increases
with increasing field.~\cite{richert1989}

A strongly disordered organic semiconductor was simulated by means of the
Gaussian disorder model~\cite{baessler1993} with a Gaussian shaped density of
states, 
\begin{equation} \label{eq:dos}
  \varrho(E) = \frac{1}{\sqrt{2 \pi \sigma^2}} \exp \left( - \frac{E^2}{2 \sigma^2} \right),
\end{equation}
where the standard deviation $\sigma$ is called the energetic disorder of the
simulated material, in conjunction with the Miller-Abrahams jump
rate~\cite{miller1960}
\begin{eqnarray} \label{eq:ma}
  \nu_{ji} &=& \nu_0 \exp(-2 \gamma r_{ji}) \times\\
 && \left\{ \begin{array}{ll}
    \exp \left(-\frac{\Delta E_{ji}}{k_{\vm{B}}T} \right), &
      \Delta E_{ji} \geq 0\\
    1, & \Delta E_{ji} < 0
    \end{array}
    \right.\nonumber
\end{eqnarray}
where $\nu_0 = 10^{13}$\,s$^{-1}$ is the attempt-to-jump frequency and
$\gamma = 5 \cdot 10^{9}$\,m$^{-1}$ is the inverse localization radius.
The first exponential function describes the tunneling of the charge and the
Boltzmann-type exponential function accounts for thermally activated jumps
upwards in energy. Hops to lower energies are not thermally activated.

A simple cubic lattice of sites with a lattice constant of 1\,nm was used. In
order to achieve a sufficient statistics for the site energies the lattice
consisted of $80 \times 40 \times 40$ sites. For a given site only the hops
from and to the 26 adjacent sites were considered. Calculations with a bigger
lattice and also further jump targets taken into account did not affect the
result.

\section{Results and discussion} \label{sec:results}

\subsection{Validity of the Einstein relation} \label{sec:einstein}

The mobility and the diffusion coefficient were calculated by the master
equation approach in conjunction with the Eqs.~(\ref{eq:mob1})
and~(\ref{eq:D}) and by the Monte Carlo approach using Eqs.~(\ref{eq:mu-mc})
and~(\ref{eq:D-mc}) respectively. The Gaussian disorder model described in
Sec.~\ref{sec:disorder} was used. In the Monte Carlo simulation, the average
and the variance of the charge carrier position has been averaged over 50.000
trajectories and the simulation time has been up to 1\,s.

\begin{figure*}
  \includegraphics[width=16.5cm]{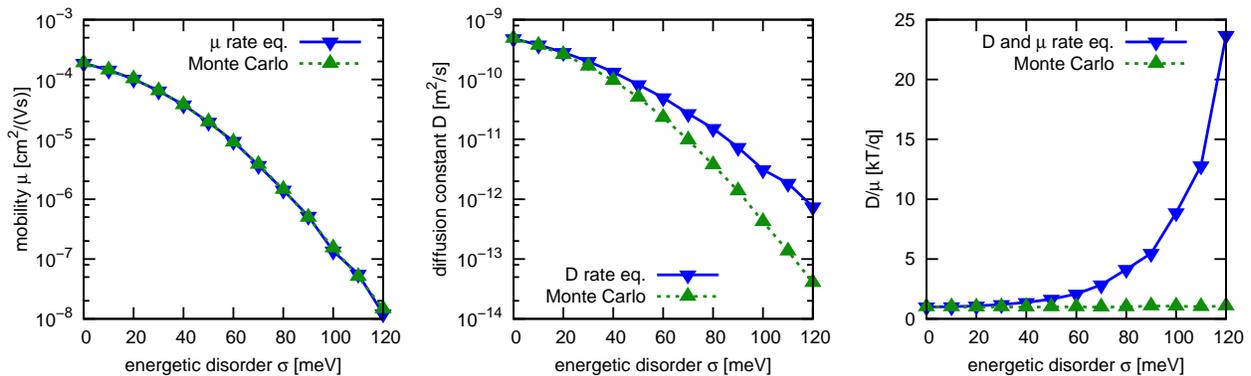}
  \caption{\label{fig:einstein_s} (Color online) Mobility, $\mu$, (left)
    diffusion coefficient, $D$, (middle) and the ratio $D/\mu$ (right) as a
    function of the energetic disorder, calculated with the rate
    equations~(\ref{eq:mob1}) and~(\ref{eq:D}) respectively, and via Monte
    Carlo simulation. The calculations were conducted at $T$ = 300\,K and $F =
    10^5$\,V/m.}
\end{figure*}

Figure~\ref{fig:einstein_s} shows the results as a function of the energetic
disorder~$\sigma$, cf.\ Eq.~(\ref{eq:dos}). The mobility varies over several
orders of magnitude and the results of Eqs.~(\ref{eq:mob1})
and~(\ref{eq:mu-mc}) match exactly. This is not the case for the diffusion
coefficient calculated with Eq.~(\ref{eq:D}) and~(\ref{eq:D-mc}). With
increasing energetic disorder, the deviations between these two approaches to
calculate $D$ increase. These deviations are not caused by the field because
for $\sigma = 0$ the results match. In order to decide which one is the right
approach, the ratio $D/\mu$ is plotted as well. One clearly sees that in the
case of Monte Carlo the Einstein relation, Eq.~(\ref{eq:einstein}), is valid,
whereas $D/\mu$ calculated with Eqs.~(\ref{eq:D}) and~(\ref{eq:mob1}) deviates
from the Einstein relation. The two mobility equations lead to the same
results. Thus, Eq.~(\ref{eq:D}) and also the frequently used
Eq.~(\ref{eq:D-simple}) provide incorrect diffusion constants for
energetically inhomogeneous materials. In any case it is advantageous to
employ the master equation in conjunction with Eq.~(\ref{eq:mob1}) to
calculate the mobility as this provides correct diffusion constants without
numerical noise and with low computational demands. 

\subsection{Angular dependence of the mobility in crystals} \label{sec:anisotropy}

\begin{figure}
  \includegraphics[width=8.5cm]{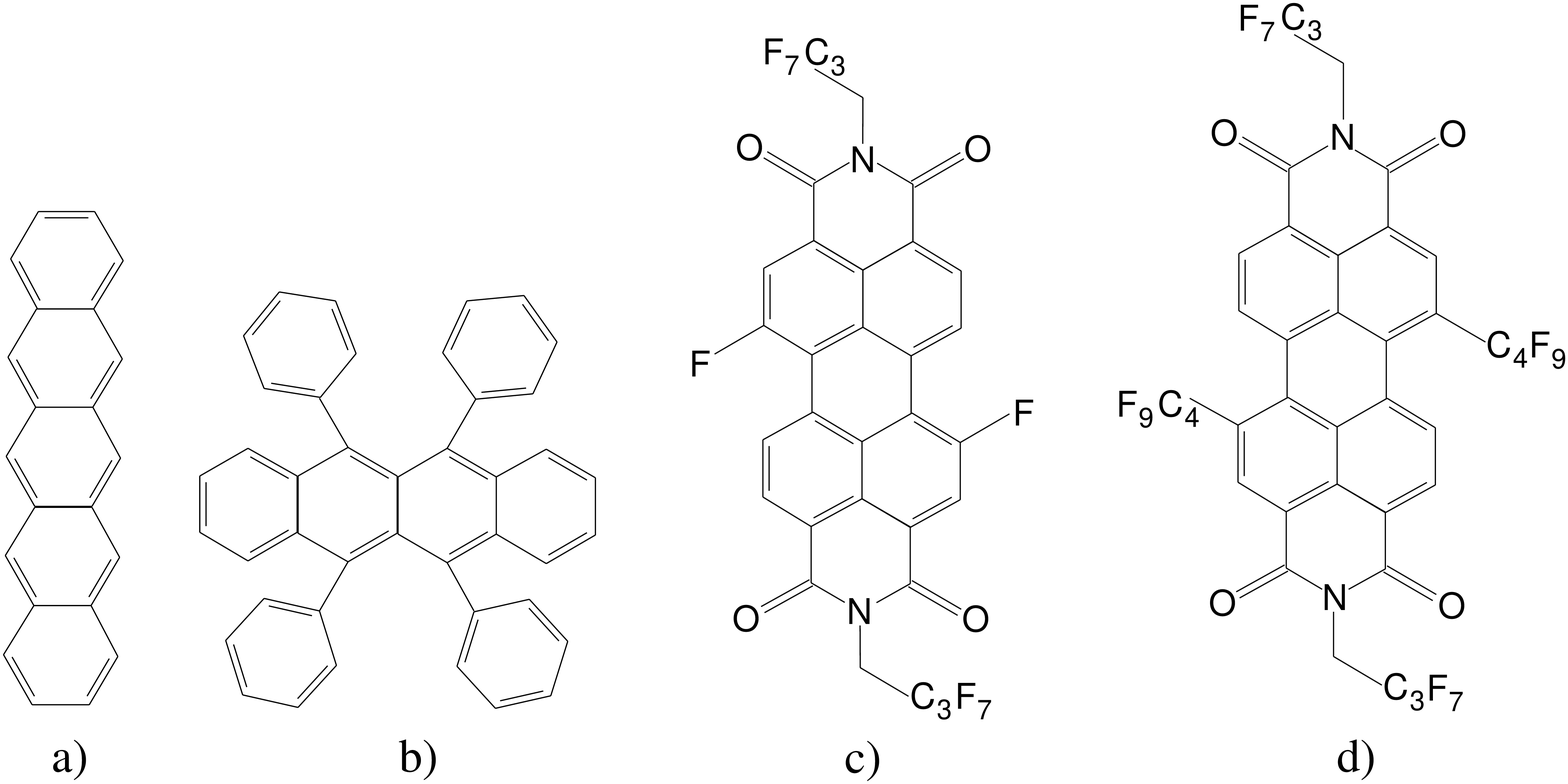}
  \caption{\label{fig:structures} The molecules investigated in this work: a)
    pentacene, b) rubrene, c) PBI-F$_2$, d) PBI-(C$_4$F$_9$)$_2$.}
\end{figure}

If not otherwise stated, the calculations have been conducted with an electric
field of $10^7$\,V/m and a temperature of 300\,K. The molecules under
investigation are depicted in Fig.~\ref{fig:structures} and the
crystallographic parameters of the corresponding crystals are listed in
Tab.~\ref{tab:crystals}.

\begin{table}
\caption{\label{tab:crystals} Lattice constants and angles for the unit cells
  of all calculated crystals.}
\begin{tabular}{l|rrr|rrr|l}
  \hline
  \hline
& a [\AA] & b [\AA] & c [\AA] & $\alpha$ [$^\circ$] & $\beta$ [$^\circ$] & $\gamma$ [$^\circ$] & Ref.\\
\hline
pentacene            &  6.27 &  7.78 & 14.53 & 76.48 &  87.68 & 84.68 & \onlinecite{mattheus2001}\\
rubrene              & 26.86 &  7.19 & 14.43 & 90.00 &  90.00 & 90.00 & \onlinecite{jurchescu2006}\\
PBI-F$_2$            & 17.46 &  5.28 & 15.28 & 90.00 & 110.90 & 90.00 & \onlinecite{schmidt2007}\\
PBI-(C$_4$F$_9$)$_2$ & 10.57 & 12.89 & 16.68 & 66.86 &  76.52 & 84.62 & \onlinecite{li2008}\\
\hline
\hline
\end{tabular}
\end{table}

\subsubsection{Pentacene} \label{sec:pentacene}

Pentacene (see Fig.~\ref{fig:structures}a) exists in several
morphologies. Here the structure described by Mattheus
{\it  et\,al.}~\cite{mattheus2001} (at 293\,K) was investigated. The unit cell 
contains two differently orientated monomers. Pentacene is known to be a hole
conductor, but for comparison, the electron transport is regarded here as
well. The reorganization energy was calculated to 92\,meV for holes and 131\,meV for
electrons. This is in good agreement with values reported before (98 and
95\,meV for holes~\cite{coropceanu2007,malagoli2004} and 132\,meV for
electrons.~\cite{coropceanu2007})

\begin{figure}
  \includegraphics[width=7cm]{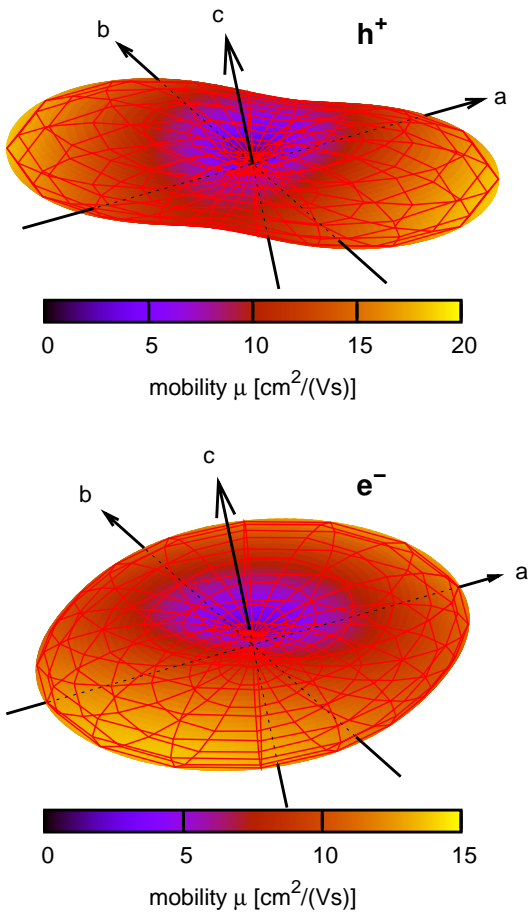}
  \caption{\label{fig:pentacene_3d} (Color online) The mobility for holes
    (top) and electrons (bottom) in the pentacene crystal for $F=10^7$\,V/m
    and $T = 300$\,K.}
\end{figure}

\begin{figure*}
  \includegraphics[width=16.5cm]{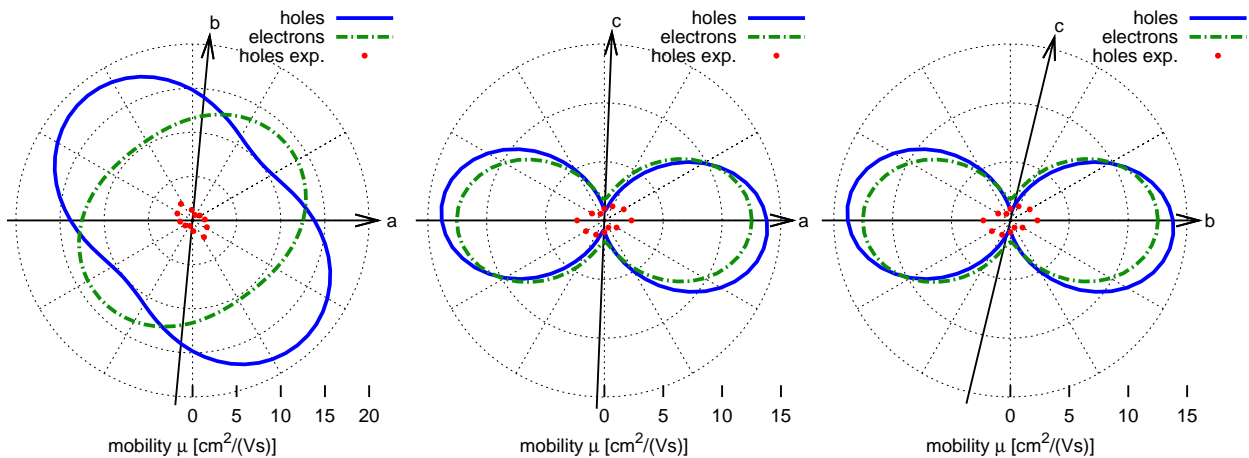}
  \caption{\label{fig:pentacene_2d} (Color online) The mobility for holes and
    electrons in the pentacene crystal in the $ab$ plane (left), $ac$ plane
    (middle) and $bc$ plane (right). The parameters are the same as in
    Fig.~\ref{fig:pentacene_3d}. For comparison some experimental
    values~\cite{lee2006} are plotted. Note that in the experiment the crystal
    orientation could not be determined~\cite{lee2006} and therefore the
    experimental data is rotated to fit best.}
\end{figure*}

\begin{figure}
  \includegraphics[width=6cm]{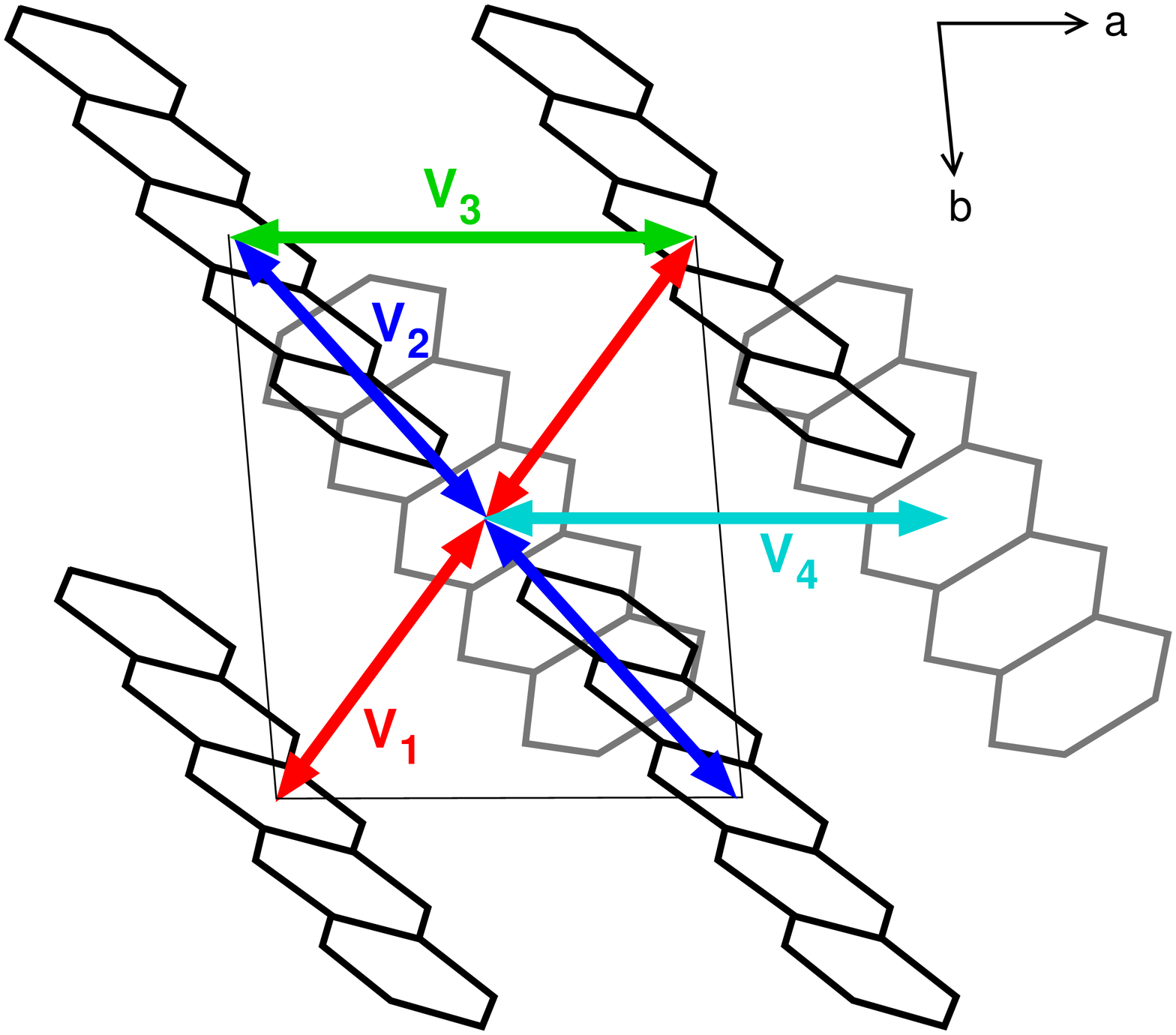}
  \caption{\label{fig:pentacene_lattice} (Color online) The most important
    hopping paths in the pentacene crystal. Direction of view is parallel to
    the $c^*$ axis.}
\end{figure}

Figure~\ref{fig:pentacene_3d} shows the mobilities of holes and electrons in
the crystal in all three dimensions. For better legibility
Fig.~\ref{fig:pentacene_2d} shows two dimensional cross sections orthogonal to
the $a^*$, $b^*$ and $c^*$ direction respectively. The magnitudes of the hole
and electron mobility are quite similar. For both types of charge
carriers the transport is almost two dimensional since the minimal
mobility, that is found in the $c^*$ direction, is very low (0.2\,cm$^2$/V\,s
for holes and 1.3\,cm$^2$/V\,s for electrons) compared with the other directions.
This can be explained by the electronic couplings. The highest ones are listed
in Tab.~\ref{tab:pentacene_Vji}. The directions of the corresponding charge
transitions are drawn in Fig.~\ref{fig:pentacene_lattice}. All of them are
coplanar in the $ab$ plane. For holes, the biggest coupling belonging to a
transition with a component in $c$ direction is one order of magnitude lower
than the lowest coupling listed in Tab.~\ref{tab:pentacene_Vji} (electrons:
about factor 5 smaller). The highest couplings for holes belong to the
transitions in [1\=10] direction, the second highest to the [110] direction.
The reverse is true for electrons. That is why the directionality of the
mobilities for holes and electrons differ in the $ab$ plane. The maximum
mobility for holes (18.5\,cm$^2$/V\,s) is found at $132^{\circ}$, the maximum
for electrons (13.7\,cm$^2$/V\,s) at $37^{\circ}$.

\begin{table}
  \caption{\label{tab:pentacene_Vji} The most important electronic couplings
    and the reorganization energy in the pentacene crystal for electrons and
    holes, cf.\ Fig.~\ref{fig:pentacene_lattice}.}
  \begin{tabular}{l|cc}
    \hline
    \hline
     & $h^+$ [meV] & $e^-$ [meV]\\
     \hline
     $V_1$     & 90.69 & 85.18\\
     $V_2$     & 55.05 & 89.66\\
     $V_3$     & 39.68 & 50.00\\
     $V_4$     & 36.62 & 47.10\\
     \hline
     $\lambda$ & 92    & 131\\
     \hline
     \hline
  \end{tabular}
\end{table}

Figure~\ref{fig:pentacene_2d} shows a comparison between the calculation and
some experimental mobility values for holes.~\cite{lee2006} Please note, that
the crystal orientation could not be determined in the
experiment.~\cite{lee2006} The measured mobility varies between 0.66 and
2.3\,cm$^2$/V\,s. This shows that the calculated maximal mobility is almost
one order of magnitude too big. However, in highly purified single crystals of
pentacene a mobility of 35\,cm$^2$/V\,s has been
measured.~\cite{jurchescu2004} It was also experimentally confirmed that the
mobility in the $ab$ plane is much larger than along the $c^*$
axis.~\cite{jurchescu2004} This is in agreement with our calculations
where the minimal mobility of about 0.2\,cm$^2$/V\,s is in $c^*$
direction. For room temperature and lower, the measurements showed a
temperature dependence of the mobility following $\mu \propto T^{-n}$ with a
positive $n$ indicating band transport.~\cite{jurchescu2004} While
this is not in accordance with the thermally activated hopping model used
here, it was also shown that above room temperature a different transport
mechanism dominates the mobility. A further reason for the overestimation of
the mobility is that the nonlocal electron-phonon
coupling\cite{hannewald2004,hannewald2004a,ortmann2010,ortmann2010a,troisi2005,troisi2006,troisi2007,stafstroem2010}
is neglected in our model. While the absolute values do not match
the measured mobilities, the qualitative dependency on the crystal direction
fits to the experimental results.

\subsubsection{Rubrene} \label{sec:rubrene}

\begin{figure}
  \includegraphics[width=8cm]{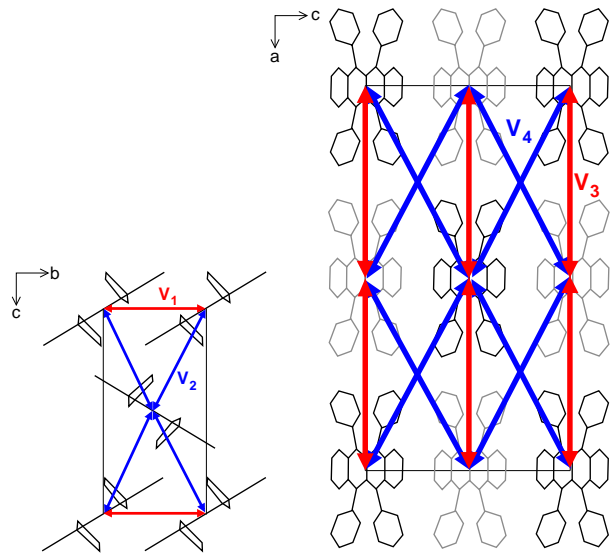}
  \caption{\label{fig:rubrene_lattice} (Color online) The most important
    hopping paths in the rubrene crystal. Direction of view is parallel to the
    $a$ axis (left) and the $b$ axis (right) respectively. The black and the
    grey monomers have a different position in $b$ direction.}
\end{figure}

Rubrene (see Fig.~\ref{fig:structures}b) is a hole conductor. It crystallizes
with four differently oriented monomers in the unit cell. The calculations
were conducted using the morphology described by Jurchescu
{\it et al.}~\cite{jurchescu2006} at 293\,K. Table~\ref{tab:rubrene_Vji} shows
the reorganization energies and the values of the four highest electronic
couplings. The couplings next in size are two orders of magnitude smaller than
the smallest coupling listed. This is in agreement with previous
calculations.~\cite{wen2009,coropceanu2007} The hopping paths corresponding to
these couplings are drawn in Fig.~\ref{fig:rubrene_lattice}. The largest
coupling ($V_1$) is between equally oriented monomers along the $b$ direction,
which is the smallest lattice constant. The second largest couplings are
between monomers which lie 
in the same plane perpendicular to the $a$ axis. $V_3$ is the coupling between
these planes and $V_4$ is the coupling between monomers in the same plane
perpendicular to the $b$ axis. 

\begin{table}
  \caption{\label{tab:rubrene_Vji} The most important electronic couplings and
    the reorganization energy in the rubrene crystal for holes and electrons,
    cf.\ Fig.~\ref{fig:rubrene_lattice}. For comparison calculated values for
    holes from Refs.~\onlinecite{wen2009} and \onlinecite{coropceanu2007} are
    shown.}
  \begin{tabular}{l|cc|cc}
    \hline
    \hline
     & $h^+$ [meV] & $e^-$ [meV] & $h^+$ [meV] \cite{wen2009} & $h^+$ [meV] \cite{coropceanu2007}\\
     \hline
     $V_1$     & 95.73 & 49.40 & 89  & 83\\
     $V_2$     & 16.38 &  5.55 & 19  & 15\\
     $V_3$     &  1.36 &  0.59 & \\
     $V_4$     &  0.24 &  0.24 & \\
     \hline
     $\lambda$ & 146   & 199   & 152 & 159\\
     \hline
     \hline
  \end{tabular}
\end{table}

\begin{figure*}
  \includegraphics[width=16.5cm]{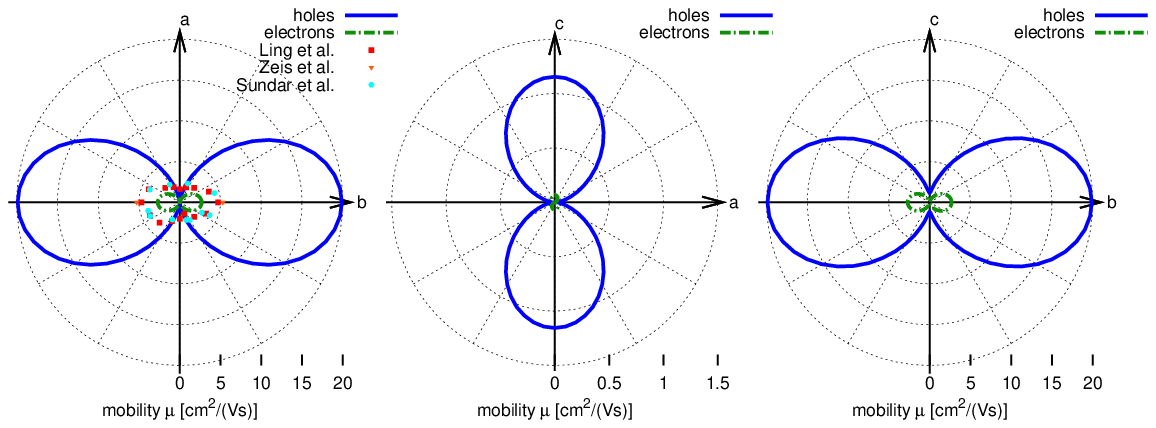}
  \caption{\label{fig:rubrene_2d} (Color online) The mobility for holes and
    electrons in the rubrene crystal in the $ba$ plane (left), $ac$ plane
    (middle) and $bc$ plane (right). The parameters are $F=10^7$\,V/m,
    $T = 300$\,K. For comparison some experimental values for hole
    mobilities~\cite{ling2007,zeis2006,sundar2004} are plotted for the
    $ba$ plane.}
\end{figure*}

\begin{figure}
  \includegraphics[width=8cm]{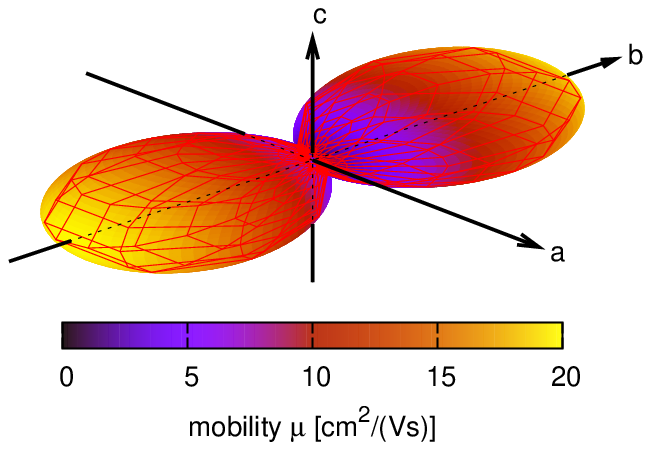}
  \caption{\label{fig:rubrene_3d} (Color online) The mobility for holes
    in the rubrene crystal in all three dimensions. The parameters are the
    same as in Fig.~\ref{fig:rubrene_2d}.}
\end{figure}

In contrast to pentacene, the electronic coupling for holes and electrons in
rubrene differs remarkably. That is why the calculated mobility for electrons
is about one order of magnitude smaller than for holes, see Fig.~\ref{fig:rubrene_2d}.
But unlike pentacene, the angular dependence of the mobility is qualitatively
the same for both types of charge carriers. For holes a three dimensional
depiction is shown in Fig.~\ref{fig:rubrene_3d}. The maximum mobility
(20\,cm$^2$/V\,s for holes and 3\,cm$^2$/V\,s for electrons) is in
$b$ direction because of the short lattice constant in that direction and the
resulting strong electronic coupling. The lowest mobility (0.03\,cm$^2$/V\,s
for holes and 0.003\,cm$^2$/V\,s for electrons) is in $a$ direction. The main
contribution to the mobility in that direction are the zig-zag jumps between
the planes perpendicular to $b$ which are marked with $V_3$  in
Fig.~\ref{fig:rubrene_lattice} and the zig-zag jumps between the planes
perpendicular to the $c$ axis marked with $V_4$. The corresponding couplings
are more than one order of magnitude smaller than the next highest
coupling~$V_2$. The zig-zag jumps corresponding to $V_2$ are the main
contribution to the mobility in $c$ direction.

Figure~\ref{fig:rubrene_2d} shows some experimental mobility values for holes
for the $ba$ plane.~\cite{sundar2004,zeis2006,ling2007} As for pentacene the
calculation overestimates the mobility. The calculated maximum mobility is
four times larger than the measured value. The mobilities for pentacene and rubrene
calculated in Ref.~\onlinecite{wen2009} with a similar approach seem to fit
better to the experiment. Yet it seems that in their calculation a wrong dwell
time of the charge carriers was used (cf.\ Sec.~\ref{sec:meq}).

The reorganization energy for rubrene is much higher than for pentacene. It
was shown that the low-frequency bending of the phenyl side-groups in rubrene
around the tetracene backbone contributes strongly to $\lambda$.~\cite{dasilva2005}
However, this bending might be impeded in the crystal and a smaller
reorganization energy would lead to an even higher mobility.

Temperature-dependent measurements in rubrene have shown a decrease of the
mobility around room temperature.~\cite{podzorov2004,boer2004} This is an
indication for band transport. However, the qualitative anisotropy
of the mobility calculated with the hopping model fits quite well to the
measurements.

\subsubsection{PBI-F$_2$} \label{sec:pbif2}

\begin{figure}
  \includegraphics[width=8.8cm]{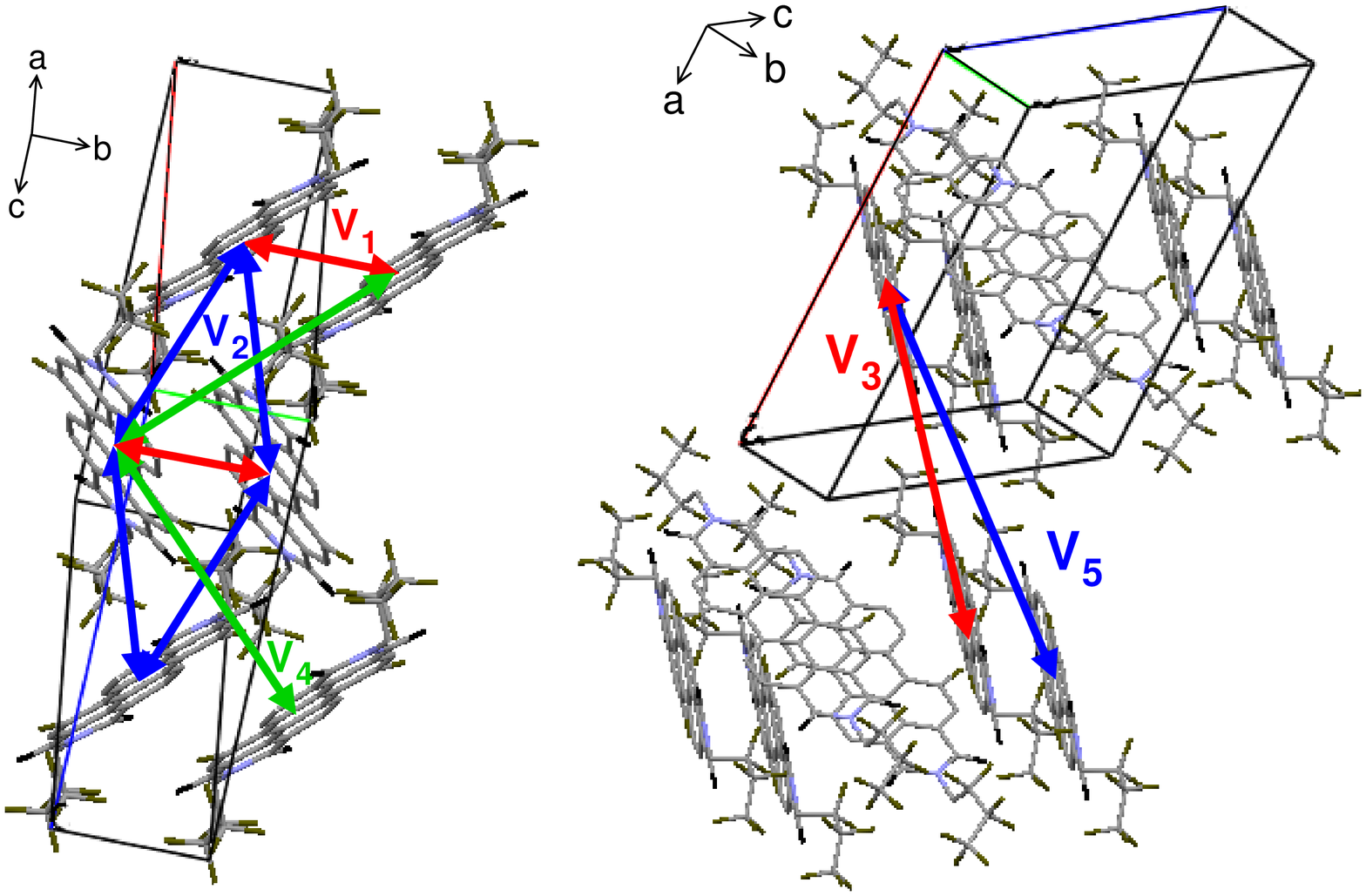}
  \caption{\label{fig:pbif2_lattice} (Color online) The most important
    hopping paths in the PBI-F$_2$ crystal.}
\end{figure}

\begin{figure}
  \includegraphics[width=7cm]{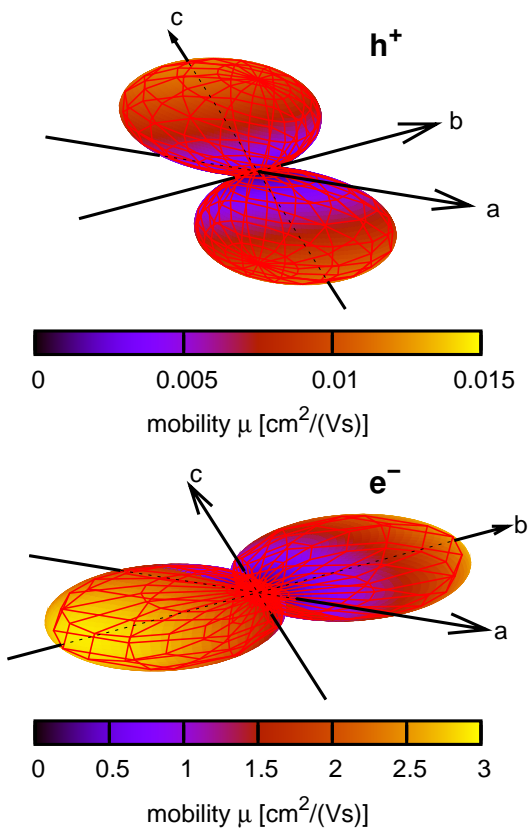}
  \caption{\label{fig:pbif2_3d} (Color online) The mobility for holes
    (top) and electrons (bottom) in the PBI-F$_2$ crystal in all three
    dimensions. The parameters are $F=10^7$\,V/m, $T = 300$\,K.}
\end{figure}

\begin{figure*}
  \includegraphics[width=16.5cm]{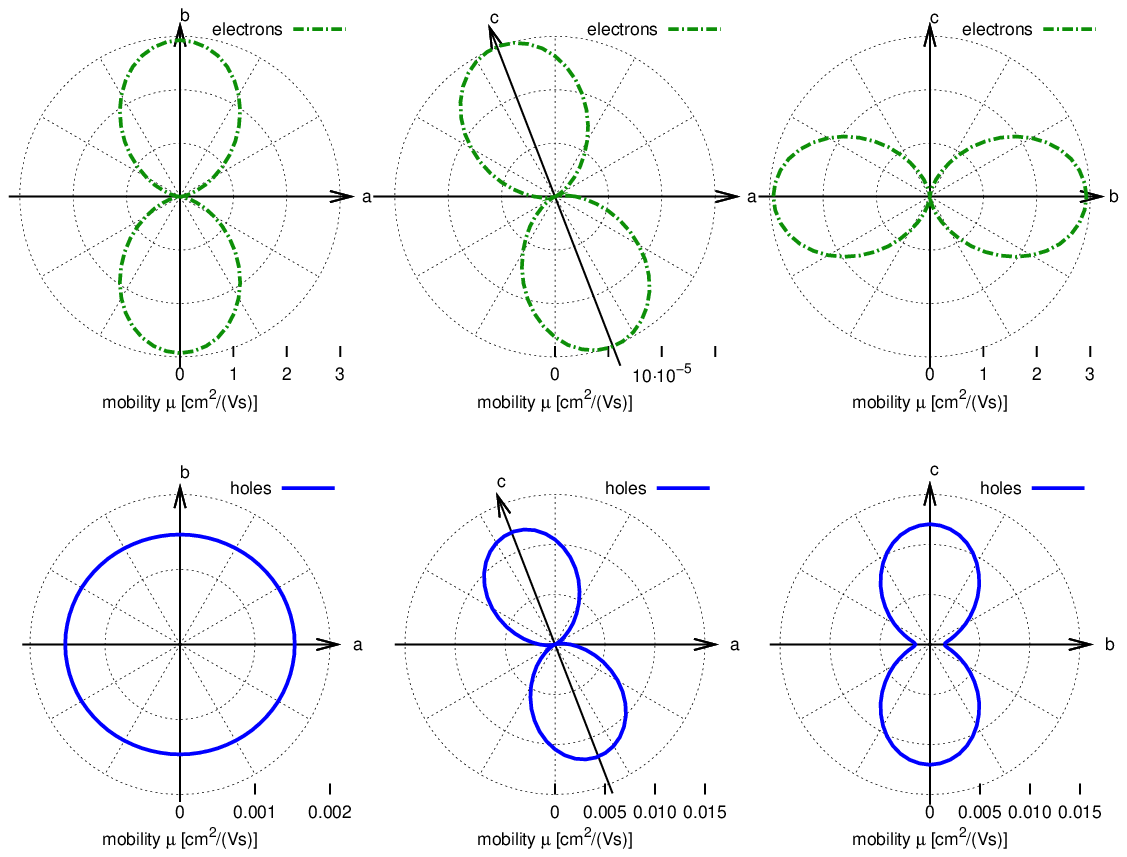}
  \caption{\label{fig:pbif2_2d} (Color online) The mobility for electrons and
    holes in the PBI-F$_2$ crystal in the $ab$ plane (left), $ac$ plane
    (middle) and $bc$ plane (right). The parameters are $F=10^7$\,V/m,
    $T = 300$\,K.}
\end{figure*}

The core-fluorinated perylene bisimide PBI-F$_2$ described by Schmidt et
al.~\cite{schmidt2007} and depicted in Fig.~\ref{fig:structures}c was analyzed.
This material is quite interesting for application since it is remarkably air
stable because of its electron-withdrawing substituents which makes the
electrons less susceptible to trapping with oxygen. The planarity of the
perylene core is only slightly distorted by the core fluorination which leads
to a torsion angle of $3^{\circ}$.~\cite{schmidt2007} It was shown that
PBI-F$_2$ has a narrower valence band and a broader conduction band than the
unsubstituted PBI, mainly due to the altered molecular
packing.~\cite{delgado2010} The unit cell contains two differently orientated
monomers. In contrast to pentacene and rubrene, PBI-F$_2$ is an electron
conductor which is caused by its high electron affinity. The electronic
couplings for electrons and holes differ remarkably. The strongest couplings
are collected in Tab.~\ref{tab:pbif2_Vji}. The couplings which are not listed are
at least one order of magnitude smaller than the smallest coupling mentioned.
The strongest coupling for electron transport is found between
monomers shifted along the $b$ direction, see
Fig.~\ref{fig:pbif2_lattice}. Note that this is
about 300 times bigger than the coupling next in size, which
is the one between two differently orientated monomers within the same unit
cell. The result is an almost one dimensional charge transport along the $b$
direction, see Fig.~\ref{fig:pbif2_3d} and~\ref{fig:pbif2_2d}. This might be
problematic for application, since the charge transport gets very sensitive to
lattice distortions, because the electron cannot easily pass at lattice
defects which cannot be avoided in real crystals.

\begin{table}
  \caption{\label{tab:pbif2_Vji} The most important electronic couplings
    and the reorganization energy in the PBI-F$_2$ crystal for electrons and
    holes, cf.\ Fig.~\ref{fig:pbif2_lattice}.}
  \begin{tabular}{l|cc|cc}
    \hline
    \hline
     & $h^+$ [meV] & $e^-$ [meV] & $h^+$ [meV] Ref. \onlinecite{delgado2010} &
     $e^-$ [meV] Ref. \onlinecite{delgado2010}\\
     \hline
     $V_1$ & 0.251 & 129.234 & 2 & 107\\
     $V_2$ & 2.398 &   0.452\\
     $V_3$ & 0.010 &   0.017\\
     $V_4$ & 0.003 &   0.004\\
     $V_5$ & 0.001 &   0.002\\
     \hline
     $\lambda$ & 213 & 303 & 215 (213) & 309 (307)\\
    \hline
    \hline
  \end{tabular}
\end{table}

\begin{figure}
  \includegraphics[width=8.5cm]{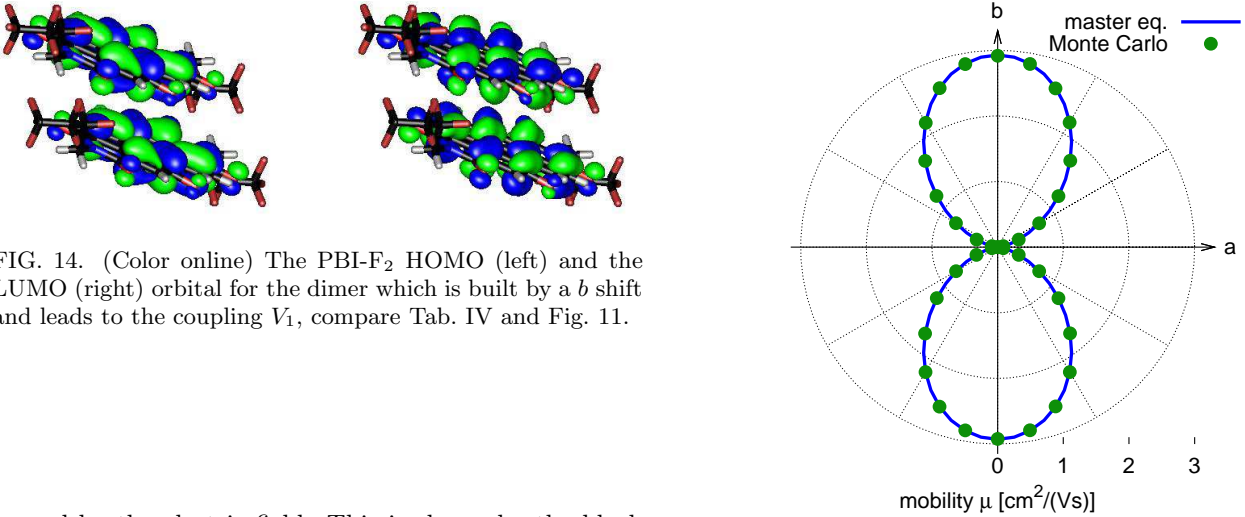}
  \caption{\label{fig:pbif2_orbitals} (Color online) The PBI-F$_2$ HOMO (left)
    and the LUMO (right) orbital for the dimer which is built by a $b$ shift
    and leads to the coupling $V_1$, compare Tab.~\ref{tab:pbif2_Vji} and
    Fig.~\ref{fig:pbif2_lattice}.}
\end{figure}

Whereas the coupling between $b$ shifted monomers is very strong for
electrons, this is surprisingly not the case for holes. Their coupling is more
than two orders of magnitude smaller than the electron coupling. This is
confirmed by other calculations.~\cite{delgado2010} The reason
can be found in the differing nodal structure of the HOMO and the LUMO orbital
for that dimer, see Fig.~\ref{fig:pbif2_orbitals}. By sliding one monomer
relative to the other along the long axis, the coupling for holes oscillates depending
on the displacement around zero,~\cite{delgado2010} because the overlap of the
two HOMO orbitals with same and different phase alternate. All the other
coupling constants do not differ significantly for the two types of charge
carriers. This sole difference in the coupling results in a maximum electron
mobility that is two orders of magnitude bigger than the maximum hole
mobility, which is achieved in $c$ direction. However, in the plane perpendicular to
$b$, the hole mobility is two orders of magnitude bigger than that of electrons,
see Fig.~\ref{fig:pbif2_2d}.

The calculated reorganization energies, 303\,meV for electrons and 213\,meV for
holes, is bigger than those for rubrene and pentacene. The values are in very
good agreement with reorganization energies calculated by Delgado
\emph{et al.}~\cite{delgado2010} (309 and 307\,meV for electrons, 215 and
213\,meV for holes).

\begin{figure}
  \includegraphics[width=6cm]{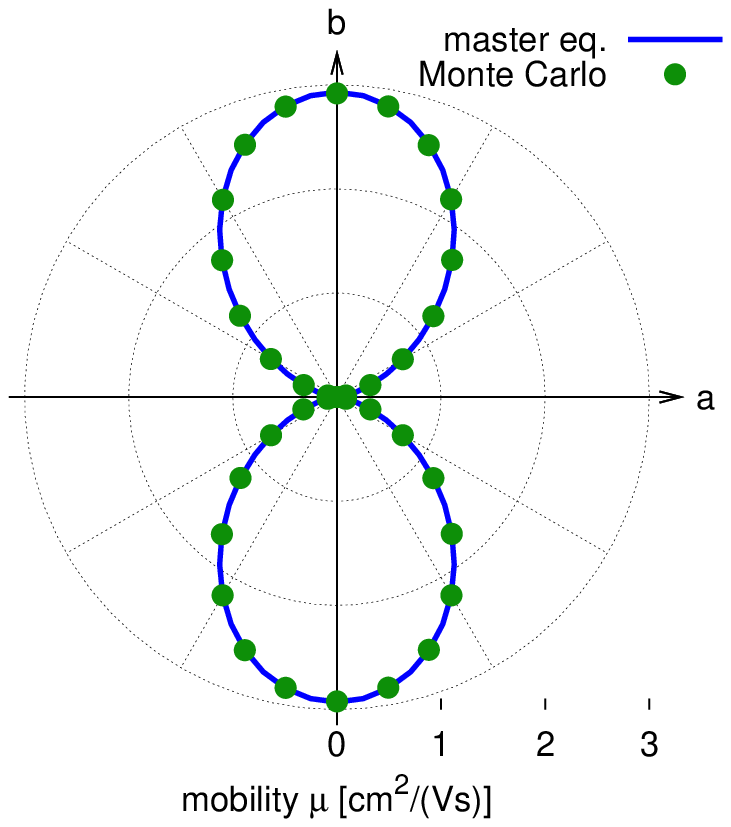}
  \caption{\label{fig:pbif2_mc} (Color online) Comparison of master equation
    and Monte Carlo results for the electron mobility in PBI-F$_2$ in the $ab$
    plane. The parameters are $F = 10^7$\,V/m, $T = 300$\,K. The two methods
    show very good agreement.}
\end{figure}

In order to test our master equation approach, some calculations were verified
with Monte Carlo calculations. The results of both methods agree very well
within the error bars of the Monte Carlo method. As an example
Fig.~\ref{fig:pbif2_mc} shows the mobility of PBI-F$_2$ in the $ab$ 
plane calculated with both approaches. The Monte Carlo simulations have run
for at least 10\,ns and have been averaged over at least 100 simulation runs,
leading to a relative average error of less than 1\,\%. For this example
the master equation approach required about 80.000 times less CPU time than
the Monte Carlo approach. Thus the master equation approach is clearly
advantageous as it is exact within the numerical accuracy of the
computer while the Monte Carlo approach contains significant and slowly
converging statistical errors.

\subsubsection{PBI-(C$_4$F$_9$)$_2$}

\begin{figure}
  \includegraphics[width=9cm]{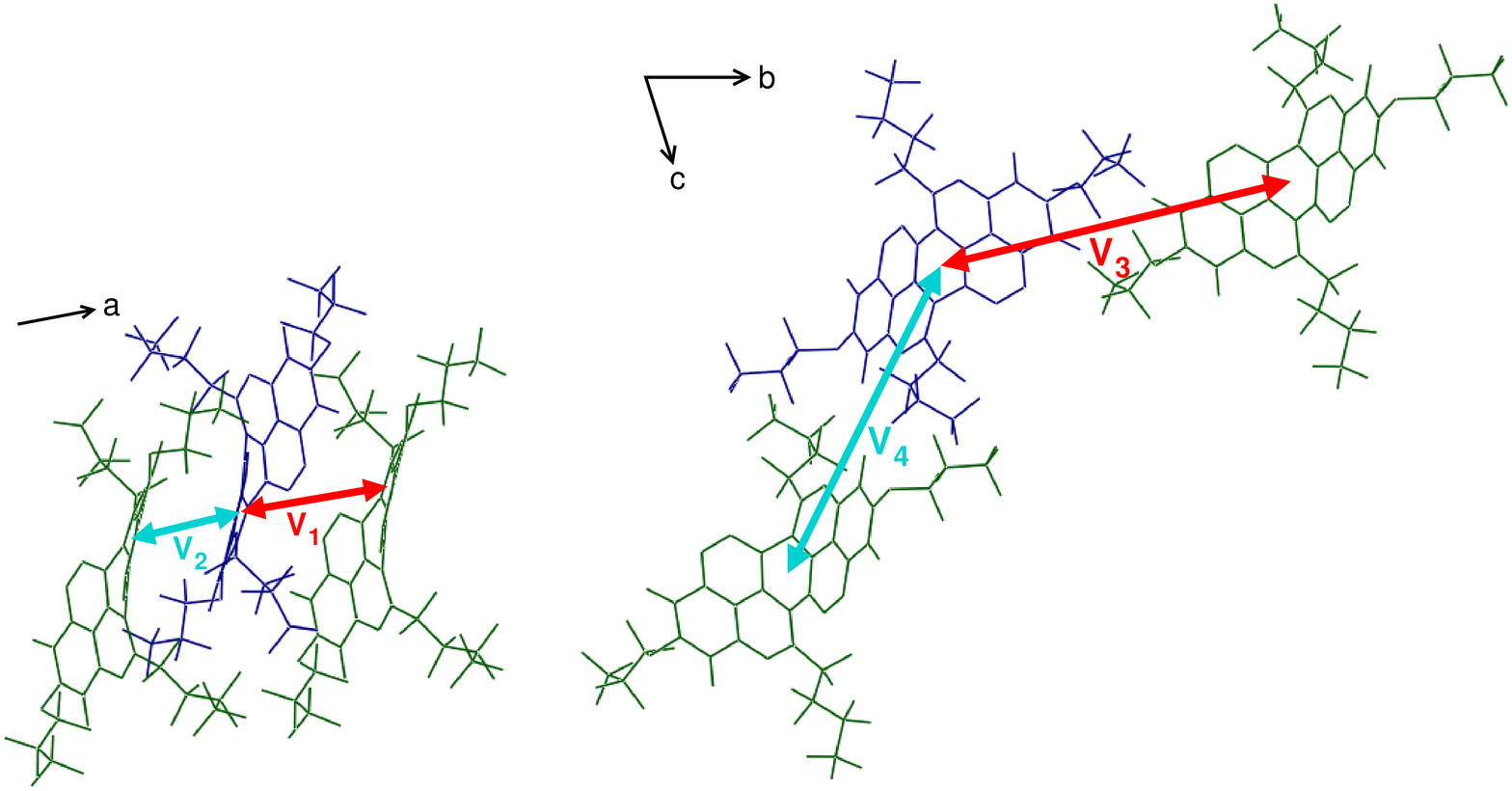}
  \caption{\label{fig:pbi-c4f9_lattice} (Color online) The most important
    hopping paths in the PBI-(C$_4$F$_9$)$_2$ crystal.}
\end{figure}

\begin{figure}
  \includegraphics[width=8.5cm]{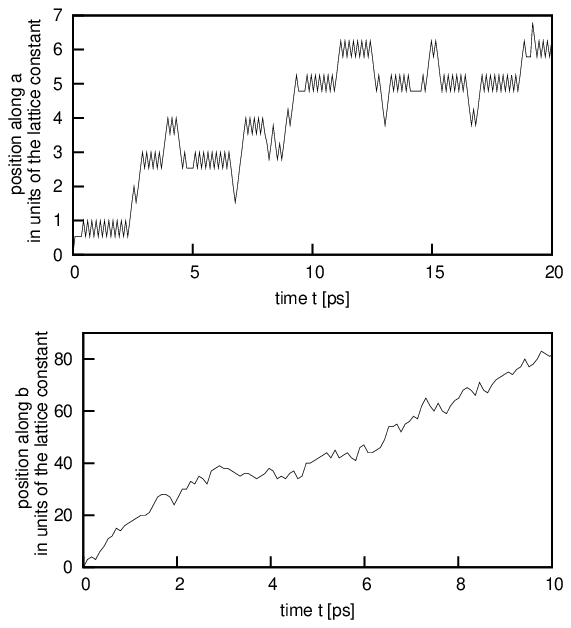}
  \caption{\label{fig:trajectories} Projection of the charge trajectory onto the
    respective direction with the highest mobility for PBI-(C$_4$F$_9$)$_2$
    ($a$~direction, top) and PBI-F$_2$ ($b$~direction, bottom). The parameters
    are $F = 10^7$\,V/m, $T = 300$\,K.}
\end{figure}

A further fluorinated perylene bisimide was investigated which was described
by Li {\it et al.}~\cite{li2008} The four most important electronic couplings are
listed in Tab.~\ref{tab:pbif2-c4f9_Vji} and depicted in
Fig.~\ref{fig:pbi-c4f9_lattice}. In contrast to the other molecules it is
striking that there is no symmetry-caused degeneration of the
electronic couplings. It is furthermore important to notice that the
intra-column couplings $V_1$ and $V_2$ along the $\pi$ stacks, which are
parallel to the $a$ axis, differ by a factor of 3. This leads to a ``trapping''
of the charge carrier between the monomers which are coupled by $V_1$ as
described in Sec.~\ref{sec:meq}: After
jumping from one monomer to the next one along $V_1$, the charge carrier is
more likely to jump back to the first monomer than to move on along $V_2$. To
illustrate this trapping a charge trajectory along the $a$ axis, simulated by
Monte Carlo, is drawn in Fig.~\ref{fig:trajectories}~(top). One clearly sees
that the charge carrier very often oscillates between two sites which lowers
the mobility of the charge along the stacks. For comparison, a charge
trajectory in PBI-F$_2$ along the high mobility axis is also depicted. No
oscillatory motions can be found there.

\begin{table}
  \caption{\label{tab:pbif2-c4f9_Vji} The most important electronic couplings
    and the reorganization energy in the PBI-(C$_4$F$_9$)$_2$ crystal for
    electrons, cf.\ Fig.~\ref{fig:pbi-c4f9_lattice}.}
  \begin{tabular}{l|c|c}
    \hline
    \hline
    & $e^-$ [meV] & $e^-$ [meV] Ref.~\cite{didonato2010}\\
    \hline
    $V_1$ & 97.7 & 95.7\\
    $V_2$ & 33.7 & 35.0\\
    $V_3$ &  2.1 &  2.2\\
    $V_4$ &  1.1 &  0.9\\
    \hline
    $\lambda$ & 339 & 360\\
    \hline
    \hline
  \end{tabular}
\end{table}

\begin{figure}
  \includegraphics[width=7cm]{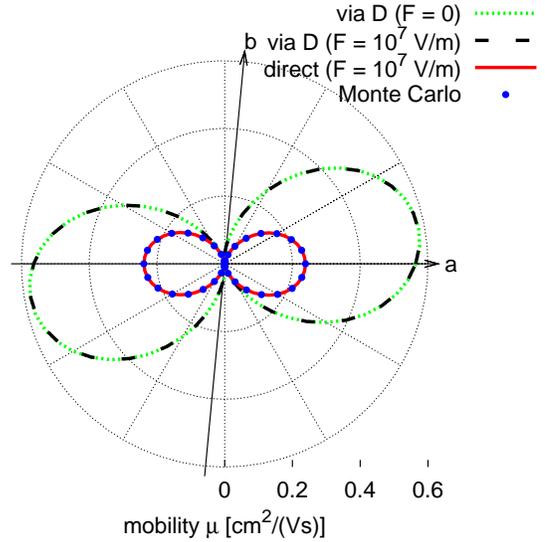}
  \caption{\label{fig:pbi-c4f9_mc} (Color online) Comparison of the mobility
    in the $ab$ plane of PBI-(C$_4$F$_9$)$_2$ calculated via the diffusion
    coefficient (Eq.~(\ref{eq:D})) and the Einstein relation
    (Eq.~(\ref{eq:einstein})) for $F = 0$ (green, dotted) and $F = 10^7$\,V/m
    (black, dashed), calculated directly (Eq.~(\ref{eq:mob1}), red, solid) and
    calculated with Monte Carlo (Eq.~(\ref{eq:mu-mc}), blue points) for $F =
    10^7$\,V/m ($T = 300$\,K in all cases).}
\end{figure}

This peculiarity of PBI-(C$_4$F$_9$)$_2$ becomes important when calculating the
mobility: Because of the ``trapping'' that is caused by these oscillations, the
mobility calculated with Eq.~(\ref{eq:D}) or~(\ref{eq:D-simple}) and the
Einstein relation~(\ref{eq:einstein}) is severely overestimated, see
Fig.~\ref{fig:pbi-c4f9_mc}. The green dotted curve is calculated without
external field with the master equation along with Eq.~(\ref{eq:D})
or~(\ref{eq:D-simple}) respectively, which is often used in literature. The
red solid curve is also obtained by the master equation but the direct
equation for the mobility, Eq.~(\ref{eq:mob1}), was applied. The maximum
mobility between these two curves differ by a factor of 2.4. Besides that, the
calculation using the diffusion coefficient and the Einstein relation even
results in a wrong angle for the maximum mobility. To prove that the result of
Eq.~(\ref{eq:mob1}) (red solid line) is the right one, Monte Carlo simulations
were conducted (blue points). The simulations ran for 10\,ns and
$\langle x \rangle$ and $\langle (x - \langle x \rangle)^2 \rangle$ were
averaged over 1000 trajectories. The relative average error was about 0.4\,\%
and the deviation of the master equation from Monte Carlo was about 0.2\,\%.
The differences in the results of Eq.~(\ref{eq:D}) or~(\ref{eq:D-simple})
and~(\ref{eq:mob1}) are not caused by the electric field. This is shown by the
black dashed line which was calculated with Eq.~(\ref{eq:D}) but with the same
field as for the red solid line. One clearly sees that the black dashed line
does not coincide with the red line but with the green line (calculated
without field) instead, proving that this approach cannot be applied.

\section{Summary and conclusions} \label{sec:conclusion}

A quantum chemical protocol for calculating the charge carrier mobilities in
organic semiconductor crystals was presented. A hopping model using Marcus
theory has been implemented by means of the master equation approach which is
more than four orders of magnitude faster than the Monte Carlo method and free
from statistical errors. In contrast to the master equation, the Monte Carlo
approach allows to simulate the transport parameters with a time dependent
framework. However, since this is a stochastic method many simulation runs are
needed in order to achieve an acceptable statistical error. Furthermore, it is
important to make sure that the stationary state is obtained within the
simulation time. This is a serious problem for disordered materials. Solving
the matrix equation~(\ref{eq:matrixequation}) describing the stationary state
instead by means of analytic numerical methods guarantees the stationary
solution.

The mobility is often calculated without external field and without the master
equation by calculating the diffusion coefficient and applying the Einstein
relation. However it can easily happen that the diffusion coefficient is
overestimated in amorphous materials and even in perfect crystals due to to a
``trapping'' of the charge between energetically similar sites. That is why it
is more appropriate to calculate the mobility by means of the master equation
from the charge drift velocity. The obtained results fit perfectly with those
of Monte Carlo simulations. It is advisable even to calculate the diffusion
coefficient out of the mobility by applying the Einstein relation,
because in the Eq.~(\ref{eq:mob1}) for the mobility, the trapping
cancels. It was shown that the Einstein relation even holds for extremely
energetically disordered materials for not too high electric fields.

The angular dependence of the mobility in pentacene, rubrene, PBI-F$_2$ and
PBI-C$_4$F$_9$  was calculated and the results were correlated with the
morphology of the crystals. The results for pentacene and rubrene show a good
qualitative agreement with experimental data. However, the absolute values of
the mobilities are strongly overestimated as he assumption of localized charge
carriers that move in a hopping process without any interaction with nonlocal
lattice vibrations is not completely adequate for organic crystals.
Nevertheless this simple model allows for qualitative transport property
predictions. It was shown that PBI-F$_2$ appears to be an almost one
dimensional n-type semiconductor.

\begin{acknowledgments}
Financial support by the Elitenetzwerk Bayern and the Deutsche
Forschungsgemeinschaft (DFG) within the framework of the Research Training
School GRK 1221 is gratefully acknowledged.
\end{acknowledgments}

\appendix

\section{} \label{sec:appendix}

Equation~(\ref{eq:D}) is valid even if an external field is applied because
in this approach the resulting drift is not caused by different jump distances
parallel or antiparallel to the field respectively since these distances
$\vec{r}_{ji}$ are fixed by the monomer positions. Instead the field
influences the jump rates $\nu_{ji}$, cf.\ Eq.~(\ref{eq:marcus}).
The drift contribution to the jump rate would have to be added to or
subtracted from the actual rate respectively. However, since $\nu_{ji}$
influences the diffusion linearly, the drift cancels when summing across all
lattice sites. In order to verify this we have computed the solution of the
time dependent master equation
\begin{equation} \label{eq:td-master}
  \frac{d}{dt} \vec p = \mathbf{N} \vec p,
\end{equation}
which reads
\begin{equation}
  \vec p(t) = \sum_i \vec c_i \vm{e}^{l_i t},
\end{equation}
with the eigenvalues $l_i$ and the respective eigenvectors $\vec c_i$.
The diffusion constant can now be calculated via 
\begin{eqnarray} \label{eq:D-1d}
  D &=& \frac{1}{2} \frac{d}{dt} \langle (x(t) - \langle x \rangle(t))^2 \rangle
        \nonumber\\
    &=& \frac{1}{2} \frac{d}{dt} \sum_i \left(p_i(t) x_i - \sum_j p_j(t) x_j\right)^2,
\end{eqnarray}
where the summation is across all sites which positions are $x_i$.
We have used the Gaussian disorder model described in
Sec.~\ref{sec:disorder} using the Miller-Abrahams hopping rate,
Eq.~(\ref{eq:ma}), for the entries of the matrix $\mathbf{N}$,
cf. Eq.~(\ref{eq:matrix}). Additionally we have used a simple biased random
walk where the mobility and the diffusion can even be calculated
analytically. Our calculations confirmed that Eq.~(\ref{eq:D}) leads to
exactly the same results as Eq.~(\ref{eq:D-1d}) as long as there is no
energetic disorder, i.e.\ $\varrho(E) = \delta(E)$, cf.\ Eq.~(\ref{eq:dos}).
The reason for the deviations in the case $\sigma \neq 0$ have already been
explained in detail in Sec.~\ref{sec:meq}.


%

\end{document}